\def\year{2026}\relax
\documentclass[letterpaper]{article} 
\usepackage{aaai2026}  
\usepackage{times}  
\usepackage{helvet}  
\usepackage{courier}  
\usepackage[hyphens]{url}  
\usepackage{graphicx} 
\usepackage{natbib}  
\usepackage{caption} 
\DeclareCaptionStyle{ruled}{labelfont=normalfont,labelsep=colon,strut=off} 
\usepackage{algorithm}
\usepackage{algorithmic}

\usepackage{xcolor}

\usepackage{newfloat}
\usepackage{listings}
\floatstyle{ruled}
\newfloat{listing}{tb}{lst}{}
\floatname{listing}{Listing}
\nocopyright
\usepackage{makecell}
\usepackage{bm}
\usepackage{enumitem}
\usepackage{multirow}
\usepackage{tabularx}
\usepackage{amssymb}
\usepackage{amsmath}
\usepackage{pifont}
\usepackage{booktabs}
\usepackage{graphicx}
\usepackage{nameref}
\usepackage{lineno}
\usepackage[subrefformat=parens,skip=0pt]{subcaption}
\usepackage[detect-all]{siunitx}
\usepackage{xspace}
\newcommand\ie{i.\,e.\xspace}
\newcommand\eg{e.\,g.\xspace}

\def\sym#1{\ifmmode^{#1}\else\(^{#1}\)\fi}

\let\namerefOld\nameref
\renewcommand{\nameref}[1]{\textit{\namerefOld{#1}}}

\newsavebox{\measurebox}

\title{Characterizing AI-Generated Misinformation on Social Media}

\author{
    Chiara Drolsbach\textsuperscript{\rm 1}, Emma Demirel\textsuperscript{\rm 1}, Nicolas Pr{\"o}llochs\textsuperscript{\rm 1}
}
\affiliations{
    \textsuperscript{\rm 1}JLU Giessen, Germany \\
    chiara.drolsbach@wi.jlug.de, emma.demirel@uni-giessen.de, nicolas.proellochs@wi.jlug.de
}

\begin{document}

\maketitle

\begin{abstract}
AI-generated misinformation (\eg, deepfakes) poses a growing threat to information integrity on social media. However, prior research has largely focused on its potential societal consequences rather than its real-world prevalence. In this study, we conduct a large-scale empirical analysis of AI-generated misinformation on the social media platform X. Specifically, we analyze a dataset comprising \num{82076} misleading posts, both AI-generated and non-AI-generated, that have been identified and flagged through X’s Community Notes platform. Our analysis yields four main findings: (i) AI-generated misinformation is more often centered on entertaining content and tends to exhibit a more positive sentiment than conventional forms of misinformation, (ii) it is perceived as less believable and less harmful than conventional misinformation, (iii) it more often originates from smaller user accounts, while authors posting such content are also associated with higher levels of partisanship and misinformation exposure,  and (iv) AI-generated misinformation is significantly more likely to go viral. Altogether, our findings highlight the unique characteristics of AI-generated misinformation on social media and offer important implications for platforms and future research. 
\end{abstract}

\section{Introduction}


\noindent
Artificial intelligence (AI) technologies are rapidly transforming the social media landscape, enabling the creation of convincing synthetic content. A prominent example is deepfakes -- AI-generated media that can realistically imitate real people’s appearance, voice, or actions \cite{Feuerriegel.2023, Hancock.2021, Groh.2024, Vaccari.2020,Sippy.2024}. These and other forms of AI-generated misinformation blur the line between authentic and fabricated content, making it increasingly difficult for users and platforms to discern what is real \cite{Goldstein.2023,Groh.2024}. As these tools become more sophisticated and accessible, they offer powerful new means for spreading misinformation \cite{Feuerriegel.2023,Goldstein.2023}. Their increased scalability, multilingualism, and multimodality further complicate detection and pose significant challenges to the defense strategies previously employed by digital platforms and users \cite{Feuerriegel.2023}. 


Despite growing concerns, AI-generated misinformation on social media remains poorly understood. Prior research warns that such content can have serious societal consequences, including the erosion of trust in media, institutions, and democratic processes \cite{Hancock.2021, Vaccari.2020, Dobber.2021,Yan.2025,Goldstein.2023}. Individuals with low media literacy may be particularly susceptible to these threats \cite{Feuerriegel.2023, Grinberg.2019}. While several studies have analyzed the ability of humans \cite{Bashardoust.2024,Kobis.2021,Groh.2024,Kreps.2022,Somoray.2023,Diel.2024,Bray.2023,Cooke.2024} and machine learning systems \cite{Zi.2020,Montserrat.2020} to detect AI-generated \mbox{content -- typically} finding that such detection is highly challenging -- there is little empirical evidence on how AI-generated misinformation actually spreads on social media and how it differs from conventional forms of misinformation. This hinders efforts to design platform defenses, guide policy, and build public resilience \cite{Feuerriegel.2023}. Our study addresses this gap by characterizing AI-generated misinformation circulating on the social media platform X (formerly Twitter).

\textbf{Research goal:}
In this paper, we conduct a large-scale empirical analysis to characterize AI-generated misinformation circulating on the social media platform X. Specifically, we address the following research questions: 
\begin{itemize}
\item \textbf{RQ~1:} \emph{How does AI-generated misinformation differ from other forms of misinformation in terms of content attributes (\eg, sentiment, topics)?}

\item \textbf{RQ~2:} \emph{How does AI-generated misinformation differ in terms of its believability and harmfulness?}
\item \textbf{RQ~3:} \emph{What are the characteristics of accounts that disseminate AI-generated misinformation?}
\item \textbf{RQ~4:} \emph{Is AI-generated misinformation more viral than other types of misinformation?}
\end{itemize}

\textbf{Data \& methods:} We analyze (and manually validate) a large dataset consisting of \num{82076} misleading posts, both AI-generated and non-AI-generated, that have been identified and flagged on X's Community Notes platform \cite{Prollochs.2022a,Twitter.2021} between January 2023 and January 2025 (\ie, during an observation period of two years). Compared to alternative approaches (\eg, manual annotation, machine learning-based identification), Community Notes offers two key advantages for identifying AI-generated misleading content: (i) it enables large-scale detection \cite{Prollochs.2022a,Pilarski.2024}, and (ii) it achieves high identification accuracy enabled by the wisdom of crowds \cite{Drolsbach.2023b,Allen.2021,Martel.2024}. To distinguish AI-generated from non-AI-generated misleading posts, we apply (and validate) a large language model (LLM) to the textual explanations provided in Community Notes. We further use (and validate) an LLM to annotate misleading posts across a wide range of content dimensions (\ie, \textit{Sentiment}, \textit{Topic}, \textit{Believability}, \textit{Harmfulness}). Based on this data, we empirically analyze how AI-generated misinformation on social media differs from traditional forms of misinformation. 

\textbf{Contributions:} To the best of our knowledge, our study is the first large-scale study characterizing AI-generated misinformation circulating on social media. Our analysis contributes the following four main findings: (i) AI-generated misinformation is more often centered on entertaining content and tends to exhibit a more positive sentiment than conventional forms of misinformation, (ii) it is perceived as less believable and less harmful than conventional misinformation, (iii) it is more likely to originate from smaller user accounts (\eg, those with fewer followers, fewer posts, and younger account ages), while authors posting such content are also associated with higher levels of partisanship and misinformation exposure, and (iv) AI-generated misinformation is significantly more likely to go viral. These findings highlight the distinct role of AI-generated content within the broader misinformation ecosystem and offer important implications for platforms, policymakers, and researchers seeking to design countermeasures against AI-generated misinformation.

\section{Background}

\textbf{Misinformation on social media: } 
Users turn to social media because these platforms offer convenient and rapid access to information, social and interactive ways of engaging with news, and feeds that can be tailored to individual interests \cite{Pew.2024}. Yet the same features that make social media appealing also introduce serious risks. Platforms provide limited oversight over the circulation of posts, enabling misinformation to spread rapidly -- often at a rate that surpasses the diffusion of accurate information \cite{Vosoughi.2018, Solovev.2022b, Prollochs.2021a, Prollochs.2023}. As a result, the proliferation of misinformation has become a defining challenge of the digital age \cite{WEF.2024}.
At the platform level, the architecture of social media amplifies these risks. Algorithmic ranking curates and prioritizes posts based on engagement signals such as reposts, likes, and replies \cite{Milli.2025,Ekstrand.2016}. Visible engagement metrics further create social proof, as highly interacted-with posts appear more credible regardless of their accuracy \cite{Avram.2020}. In addition, personalized feeds and self-selected follower networks can reinforce ideological echo chambers in which users predominantly encounter like-minded perspectives \cite{Barbera.2015}, ultimately reducing exposure to accurate information \cite{Pennycook.2018}. 
At the individual level, users often accept information without verifying its accuracy \cite{Prollochs.2023,Vo.2018, Pennycook.2021}, either because they pay limited attention to whether the content is reliable \cite{Ceylan.2023} or because they are unwilling to invest additional effort \cite{Geeng.2020}. Moreover, misleading content is often written to intentionally deceive, which complicates the detection of misinformation \cite{Wu.2019}. Taken together, these platform- and individual-level mechanisms enable misinformation to thrive on social media. In response, researchers, governments, and regulation authorities urge social media providers (\eg, X/Twitter, Facebook) to develop effective countermeasures to counteract the spread of misinformation on their platforms \cite{Lazer.2018,Calo.2021,Donovan.2020,Feuerriegel.2023,Kozyreva.2022}.

\textbf{AI-generated misinformation: }
The manipulation of media content has long been a concern; however, the rapid advancements of generative AI tools have significantly lowered the barriers to creating sophisticated AI-generated content \cite{Westerlund.2019,Feuerriegel.2023}. AI-generated content (\eg, deepfakes) -- artificially generated videos, images, and audio designed to mimic real individuals -- has become increasingly prevalent on social media, raising critical concerns about their role in spreading misinformation \cite{Hancock.2021,Groh.2024,Vaccari.2020}. This emerging form of digital misinformation poses significant challenges to the defense strategies previously employed by digital platforms and users \cite{Goldstein.2023,Feuerriegel.2023}. Additionally, the increased scalability, multilingualism, and multimodality of AI-generated content further complicate its detection and mitigation \cite{Feuerriegel.2023,Timmerman.2023}.
Research on AI-generated misinformation is still in its early stages, with most existing studies focusing on their potential societal impacts, such as the erosion of trust in media and democratic institutions \cite{Dobber.2021,Hancock.2021,Vaccari.2020,Goldstein.2023}. Despite growing concern, there is limited empirical evidence on the actual prevalence of AI-generated misinformation on social media and how it spreads relative to other forms of misinformation. Although AI-generated misinformation is often portrayed as a particularly dangerous type of digital misinformation \cite{Hancock.2021}, it remains unclear what specifically distinguishes them from traditional misinformation in terms of its content characteristics, reach, and user engagement.
A separate body of research has focused on detecting AI-generated misinformation, typically through one of two approaches: (1) human-centered methods \cite{Groh.2024}, and (2) automated, machine learning-based methods \cite{Zhou.2023,Montserrat.2020}. Human-centered methods, which are dependent on individual judgments, often struggle to accurately identify AI-generated content \cite{Groh.2024,Somoray.2023,Diel.2024,Bray.2023}.
Further, they lack scalability, which makes them ineffective for large volumes of media \cite{Groh.2024}. Yet, evidence from studies on traditional misinformation shows that crowd-based approaches, where multiple judgments are aggregated, can achieve accuracy levels comparable to those of experts and overcome scalability issues \cite{Allen.2021,Martel.2024,Drolsbach.2023b,He.2025, Pilarski.2026}.
In contrast, automated detection approaches, \eg, using AI models to analyze inconsistencies in facial expressions, lighting, and audio-visual mismatches, offer a scalable solution. However, they often struggle with accuracy and robustness, leading to false positives or undetected deepfakes \cite{Almars.2021,Feuerriegel.2023,Zhou.2023}. To enable comprehensive and reliable empirical analysis of AI-generated misinformation circulating on social media, there is thus a need for complementary detection strategies that are both scalable and accurate.

\textbf{Our work: } 
In this study, we characterize AI-generated misinformation that has been identified on X's Community Notes platform, \ie, via the wisdom of crowds. Compared to other identification strategies, this has two key advantages: (i) it overcomes the scalability limitations of human-centered methods \cite{Groh.2024}, and (ii) it addresses challenges with low accuracy of AI-based approaches for the detection of AI-generated content \cite{Almars.2021,Feuerriegel.2023,Zhou.2023}. This unique data source enables us to empirically analyze what distinguishes AI-generated misinformation from traditional forms of misinformation in terms of reach, user engagement, and content characteristics. Addressing these questions is a crucial first step for developing effective countermeasures against AI-generated misinformation.

\section{Data and Methods}

\subsection{Data Source}
\begin{figure*}[t]
\centering
\includegraphics[width=0.8\linewidth]{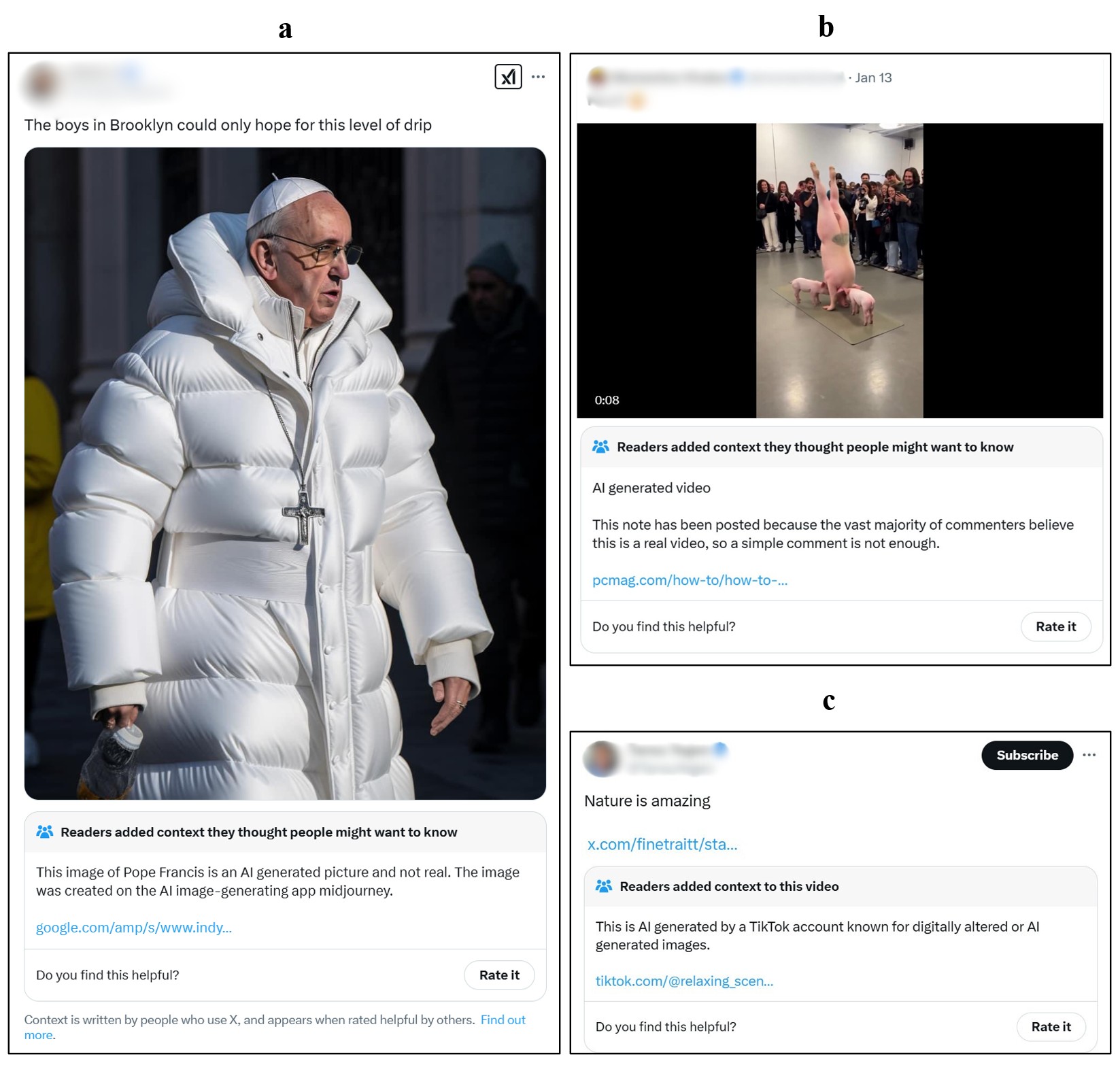}
\caption{Examples of posts on X that share a deepfake in form of an (\textbf{a}) image, (\textbf{b}) video, or (\textbf{c}) no media (here: URL to media content) and the corresponding Community Notes.}
\label{fig:example}
\end{figure*}

To address our research questions, we analyze a comprehensive dataset of crowd-annotated social media posts flagged as misleading on X's Community Notes between January 2023 and January 2025, \ie, for an observation period of two years.\footnote{Available via \url{https://communitynotes.x.com/guide/en/under-the-hood/download-data}} X's Community Notes is a crowdsourced fact-checking system that enables users to add context to potentially misleading posts \cite{Prollochs.2022a,Twitter.2021}. By incorporating contributions from diverse perspectives, the system strives to provide balanced and informative notes \cite{Wojcik.2022} to help users assess content accuracy \cite{Drolsbach.2024, Bobek.2026}. Each Community Note consists of the assigned label (\ie, whether the post is considered misleading) and a textual description that explains why a post is misleading (\eg, because it is a deepfake; see examples in Fig.~\ref{fig:example}). Further, Community Notes features a rating mechanism, where other users can rate the helpfulness of each note \cite{Solovev.2025}. A note becomes visible on X for all users if (and only if) it reaches a certain helpfulness rating, reflecting agreement across diverse perspectives rather than relying solely on simple majority approval. Community fact-checks identified as helpful on the Community Notes platform have been shown to be accurate \cite{Drolsbach.2023b} and trustworthy \cite{Drolsbach.2024}.

For our data collection, we filtered all Community Notes rated as helpful, resulting in a dataset of \num{94343} for \num{82076} source posts. . 
To retrieve additional metadata about the fact-checked posts, we mapped the referenced \textit{postID} to the original source post via the X Research API. This includes key attributes such as the number of reposts, likes, and impressions, the type of attached media, as well as details about the author's profile, including follower count, followee count, total post count, account age, and verification status. We further downloaded all attached media (\ie, images, and videos).
Finally, we estimated the authors' political partisanship and misinformation exposure on social media using the method proposed by \citet{Mosleh.2022}. Partisanship scores range from $-1$ (Democrat) to $+1$ (Republican) and are based on the number of Democratic and Republican public figures followed by each user. The misinformation exposure score ($[0,1]$) is derived from the proportion of followed public figures that have been rated false by PolitiFact.\footnote{We use the method available at \url{https://github.com/mmosleh/minfo-exposure}. Due to X API restrictions, partisanship and misinformation expose could only be calculated for authors of \num{27985} posts in our dataset.}

\subsection{Identification of AI-generated Misinformation}
\label{sec:AI_identification}

To distinguish AI-generated content from other misleading posts, we implemented an LLM-based identification approach \cite{Feuerriegel.2025, Chuai.2025b}. Specifically, we used the \texttt{27B} parameter version of \texttt{Gemma 3} \citep{Gemma.2025} to determine whether a Community Note refers to AI-generated content. Rather than classifying the posts directly, we applied the model to the textual explanation in Community Notes, which frequently state whether content is AI-generated (see examples in Fig.~\ref{fig:example}). We used these references as indicators that the corresponding post contains AI-generated content. Here, AI-generated content refers to any type of media (\eg, text, image, audio, or video) that is created using artificial intelligence technologies, including outputs from machine learning models such as LLMs, image generators, and speech synthesis systems. 
For each case, the model returned a binary label (``Yes'' or ``No'') indicating whether the note referred to AI-generated content, together with a confidence score (``Low'', ``Medium'', or ``High''; see SI, \emph{Sec.}~\nameref{sec:identification_prompt} for the full prompt). Because posts can receive multiple Community Notes, we first annotated each note individually ($N =$ \num{94343}) and then aggregated the results at the post level ($N =$ \num{82076}). A post was classified as AI-generated if at least one associated helpful Community Note referred to AI-generated content. 

Compared to alternative strategies (\eg, keyword-based heuristics, purely content-based methods), our approach to identify AI-generated misleading posts via Community Notes offers two main advantages: (i) it leverages crowd-sourced fact-checking explanations (\ie, Community Notes), which contain contextual insights not evident from the media content alone, and (ii) it achieves comparatively high accuracy (see \emph{Sec.}~\nameref{sec:validation}).

\subsection{Annotation of Post Characteristics}
\label{sec:annotation}

We further annotated all source posts in the \mbox{dataset ($N =$ \num{82076})} with a wide variety of content characteristics. For this task, we again used the \texttt{27B} parameter version of \texttt{Gemma 3} \citep{Gemma.2025}, leveraging both the textual content of the posts and any attached media (images or video snapshots).
The model was also provided with the information that all posts had been flagged as misleading by a community-based fact-checking system. The annotation focused on four core dimensions: \textit{Sentiment}, \textit{Topic}, \textit{Harmfulness}, and \textit{Believability} (see SI, \emph{Sec.}~\nameref{sec:example_annotation} for examples).

\textit{Sentiment} \cite{Feuerriegel.2025} refers to the emotional tone or attitude expressed in the post, categorized as ``Positive'', ``Neutral'', or ``Negative''. The \textit{Topic} dimension \cite{Feuerriegel.2025} required the LLM to identify the main subject matter as one of the following: ``Politics'', ``Technology'', ``Health'', ``Crime'', ``Business'', ``Entertainment'', ``Sports'', ``Education'', ``Satire'', and ``Other''. \textit{Harmfulness} \cite{Drolsbach.2023a} assessed the potential of the content to cause real-world damage, including emotional distress, physical harm, or societal disruption, particularly if the misinformation were believed and acted upon (``Low'', ``Medium'', or ``High''). \textit{Believability} \cite{Drolsbach.2023a} was defined as the degree to which a post could plausibly be accepted as true by a general audience (``Low'', ``Medium'', or ``High''). These definitions were embedded directly in the system prompt to ensure consistent and interpretable outputs. The full prompt is available in the SI, \emph{Sec.}~\nameref{sec:annotation_prompt}.

\subsection{Validation}
\label{sec:validation}
We validate the LLM-generated annotations using pairwise comparisons conducted by three trained human coders. Pairwise evaluations are particularly well suited for subjective or difficult-to-define constructs such as \textit{AI-generated}, \textit{Believability}, and \textit{Harmfulness} \citep{Thurstone.1927, Thurstone.1954}, as comparative judgments have been shown to yield more reliable and efficient assessments than absolute classifications \citep{Thwaites.2024}. For the \textit{AI-generated} label, coders evaluated \num{100} pairs of posts and selected the post that appeared more likely to be AI-generated. For each remaining characteristic, we randomly sampled \num{100} post pairs representing opposing labels (\ie negative vs. positive \textit{Sentiment}, low vs. high \textit{Believability}, and low vs. high \textit{Harmfulness}). For \textit{Topic}, coders evaluated $100$ post pairs for each topic category (\ie, \textit{Technology}, \textit{Sports}, \textit{Business}, \textit{Health}, \textit{Crime}, \textit{Satire}, \textit{Entertainment}, \textit{Politics}, and \textit{Education}). Each pair consisted of one post annotated with the respective topic and one post without that label. In total, the validation study comprised \num{1300} post pairs (\ie, \num{2600} posts). Across all evaluated characteristics (see SI, Tab.~\ref{tab:user_study_pair}), human judgments showed strong agreement with the LLM-generated labels, with average accuracies ranging from $0.83$ to $0.99$ and Fleiss’ $\kappa$ values between $0.64$ and $0.96$, indicating substantial to almost perfect agreement among human coders \cite{Landis.1977}. These results indicate high annotation reliability and support the validity of the automated labeling procedure.

\section{Empirical Analysis}

\subsection{Content Characteristics (RQ~1)}

\begin{figure}[t]
	\captionsetup{position=top}
	\captionsetup{belowskip=1pt}
	\captionsetup[subfloat]{textfont={sf,normalsize}, skip=2pt, singlelinecheck=false, labelformat=simple,labelfont=bf,justification=centering}
	\centering
	\includegraphics[width=0.9\linewidth]{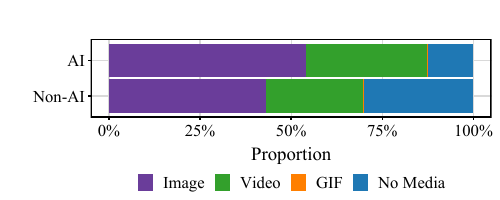}
	\caption{Distribution of media types/modalities in AI-generated vs. non-AI-generated misleading posts. \mbox{$\bm{N}=$ \num{82076}}.}
	\label{fig:media_types}
\end{figure}

Our final dataset (see SI, Tab.~\ref{tab:data_summary} for summary statistics) includes \num{82076} posts across more than $50$ languages, with \num{6024} posts (\ie, \SI{7.33}{\percent}) identified as containing AI-generated content, with the vast majority of classifications being assigned with high confidence (\SI{99.03}{\percent}). English dominates the dataset, accounting for half of all posts (\SI{49.20}{\percent}), followed by Japanese (\SI{12.35}{\percent}), \mbox{Spanish (\SI{11.04}{\percent})}, French (\SI{8.03}{\percent}), Portuguese (\SI{6.11}{\percent}), and German (\SI{2.66}{\percent}). A long tail of low-frequency languages reflects the dataset's global scope, though many languages are represented by only a handful of posts. As shown in Figure \ref{fig:media_types}, AI-generated misleading posts are considerably more visual than non-AI-generated misleading posts, containing more images (\SI{54.17}{\percent} vs. \SI{43.10}{\percent}), more videos (\SI{33.10}{\percent} vs. \SI{26.48}{\percent}), and fewer posts without media (\SI{12.63}{\percent} vs. \SI{30.16}{\percent}), whereas animated GIFs are rare in both groups (\SI{0.10}{\percent} vs. \SI{0.26}{\percent}). Compared to other forms of misinformation, AI-generated misleading posts are $1.25$ times as likely to contain media elements ($RR = 1.25$,\, $p<0.001$).

\begin{figure}[t]
	\captionsetup{position=top}
	\captionsetup{belowskip=1pt}
	\captionsetup[subfloat]{textfont={sf,normalsize}, skip=2pt, singlelinecheck=false, labelformat=simple,labelfont=bf,justification=centering}
	\centering
	\includegraphics[width=0.9\linewidth]{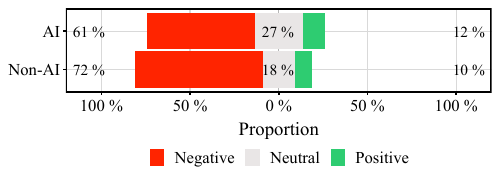}
	\caption{Distribution of sentiment in AI-generated vs. non-AI-generated misleading posts. $\bm{N}=$ \num{82076}.}
	\label{fig:sentiment}
\end{figure}

\textbf{Sentiment: } We applied an LLM-based \textit{Sentiment} classification to the dataset that incorporates both the textual content of each post \textit{and} any attached media. Figure \ref{fig:sentiment} presents the distribution of posts classified as negative, neutral, or positive. The results indicate that a higher proportion of AI-generated posts exhibit positive \textit{Sentiment} compared to non-AI-generated posts (\SI{12.37}{\percent} vs. \SI{10.07}{\percent}). In contrast, negative \textit{Sentiment} is more prevalent among non-AI-generated posts (\SI{72.25}{\percent} vs. \SI{60.56}{\percent}). Further, Pearson’s chi-squared test revealed a statistically significant association between AI generation status and \textit{Sentiment} \mbox{classification ($\chi^2 = 403.5,\, p<0.001$)}.
This indicates that the distribution of \textit{Sentiment} categories differs systematically between AI-generated and non-AI-generated posts and suggests that AI-generated misinformation may be more likely to adopt a humorous or entertaining tone.

\begin{figure}[t]
	\captionsetup{position=top}
	\captionsetup{belowskip=1pt}
	\captionsetup[subfloat]{textfont={sf,normalsize}, skip=2pt, singlelinecheck=false, labelformat=simple,labelfont=bf,justification=centering}
	\centering
	\includegraphics[width=0.9\linewidth]{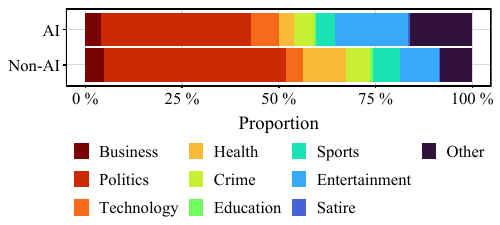}
	\caption{Distribution of topics in AI-generated vs. non-AI-generated misleading posts. $\bm{N}=$ \num{82076}.}
	\label{fig:topics}
\end{figure}

\textbf{Topics: } Based on the LLM-based \textit{Topic} classification (including both textual content \& media), we observe significant differences in the distribution of specific topics between AI-generated and non-AI-generated content (see Fig. \ref{fig:topics}), as indicated by a Pearson’s chi-squared test \mbox{($\chi^2 = 1323.1,\,p < 0.001$)}. Specifically, \SI{18.73}{\percent} of AI-generated posts are associated with the topic \textit{Entertainment}, while only \SI{9.89}{\percent} of non-AI-generated posts focus on this topic. This suggests that AI-generated content is more likely to focus on lighter, more engaging topics. In contrast, the share of posts related to the topic \textit{Health} is higher among non-AI-generated content \mbox{(\SI{11.22}{\percent} vs. \SI{3.80}{\percent})}, whereas about \SI{46.94}{\percent} of non-AI-generated posts focus on \textit{Politics}, compared to \SI{38.78}{\percent} for AI-generated posts, indicating a potential difference in the focus areas of human versus AI creators. For all other topics, the distribution remains largely similar between the two groups \mbox{(\ie, the differences are $\leq$ \SI{3.00}{\percent}}).

\subsection{Harmfulness \& Believability (RQ~2)} 

As AI models become increasingly capable  of mimicking human communication, their ability to produce more credible and potentially more harmful misinformation grows \cite{Feuerriegel.2023}. Understanding characteristics such as \textit{Believability} and \textit{Harmfulness} is therefore critical for assessing the risks associated with AI-generated content and for informing effective detection and mitigation strategies \cite{Drolsbach.2023a}. To address this, we analyze the LLM-based annotation of post \textit{Believability} and \textit{Harmfulness}, comparing AI-generated and non-AI-generated misinformation.

We find that the majority of posts are perceived as having medium levels of \textit{Harmfulness}, accounting for \SI{57.17}{\percent} of AI-generated and \SI{66.92}{\percent} of non-AI-generated posts. A substantial share of posts is also classified as low in \textit{Harmfulness} (\SI{38.13}{\percent} for AI-generated and \SI{27.61}{\percent} for non-AI-generated posts), whereas only a small fraction is perceived as highly harmful (both around \SI{5.00}{\percent}). In terms of \textit{Believability}, most posts are perceived as having low \textit{Believability}, with \SI{76.83}{\percent} of AI-generated and \SI{61.50}{\percent} of non-AI-generated posts falling into this category. Medium \textit{Believability} is more common among non-AI-generated posts (\SI{32.64}{\percent} versus \SI{16.87}{\percent}), while high believability is equally rare across both content types (both around \SI{6.00}{\percent}). Overall, the distributions of both \textit{Believability} \mbox{($\chi^2 = 652.4, p<0.001$)} and \textit{Harmfulness} \mbox{($\chi^2 = 303.6, p<0.001$)} differ significantly between AI-generated and non-AI-generated misinformation.

\begin{figure}[t]
    \captionsetup{position=top}
    \captionsetup{belowskip=1pt}
    \captionsetup[subfloat]{textfont={sf,normalsize}, skip=2pt, singlelinecheck=false, labelformat=simple, labelfont=bf, justification=centering}
    \centering

    \subfloat[]{%
        \includegraphics[width=0.45\linewidth]{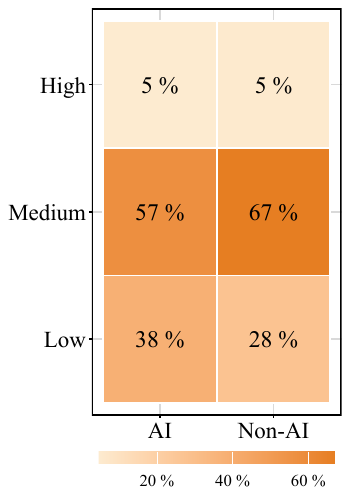}
    }\hfill
    \subfloat[]{%
        \includegraphics[width=0.45\linewidth]{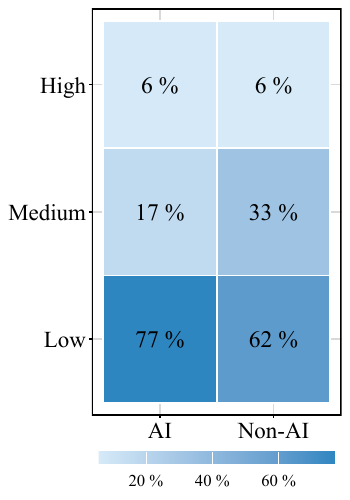}
    }
    \caption{{Harmfulness} (\textbf{a}) and {believability} (\textbf{b}) of AI-generated vs. non-AI-generated misleading posts. \mbox{$\bm{N}=$ \num{82076}.}}
    \label{fig:harm_bel}
\end{figure}

\subsection{Author Characteristics (RQ~3)}

\begin{figure}[t]
    \captionsetup{position=top}
    \captionsetup{belowskip=1pt}
    \captionsetup[subfloat]{textfont={sf,normalsize}, skip=2pt, singlelinecheck=false, labelformat=simple, labelfont=bf, justification=centering}
    \centering

    \subfloat[]{%
        \includegraphics[width=0.45\linewidth]{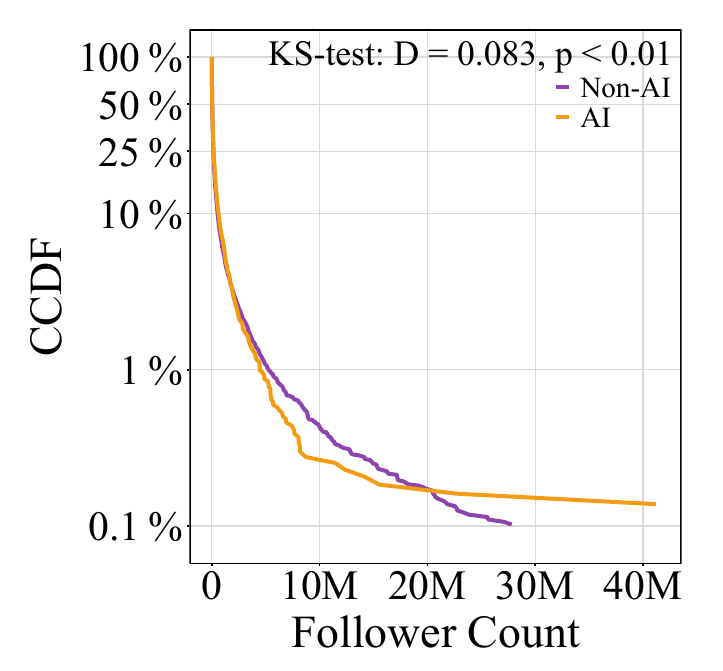}
    }\hfill
    \subfloat[]{%
        \includegraphics[width=0.45\linewidth]{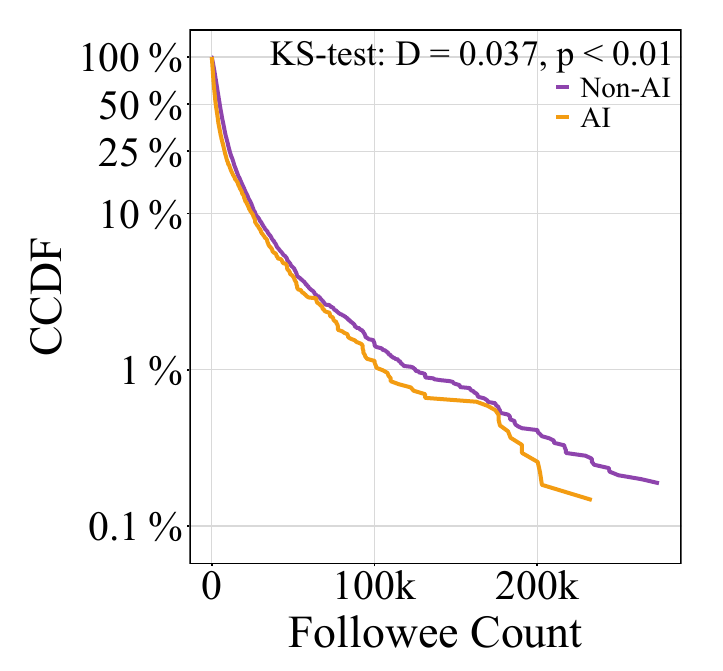}
    }

    \subfloat[]{%
        \includegraphics[width=0.45\linewidth]{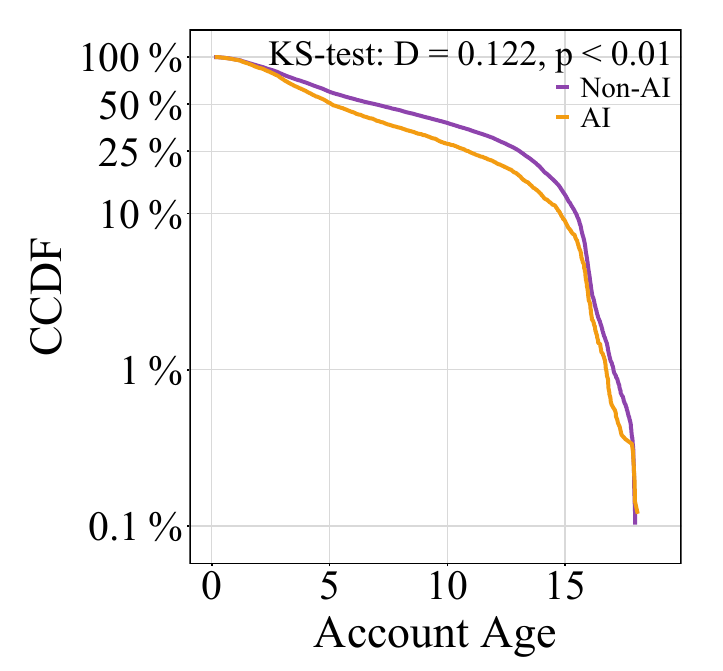}
    }\hfill
    \subfloat[]{%
        \includegraphics[width=0.45\linewidth]{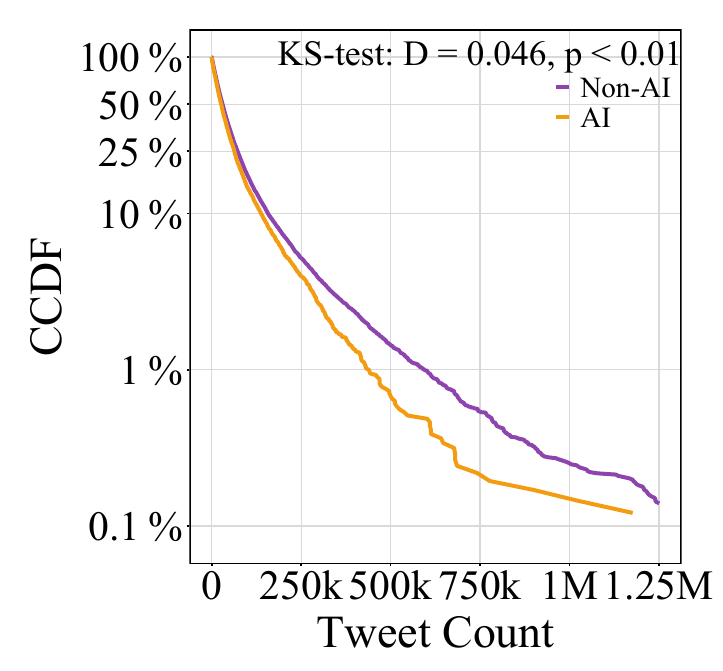}
    }

    \caption{Complementary cumulative distribution functions (CCDFs) for AI vs. non-AI-generated misleading posts, shown separately for follower counts (\textbf{a}), followee counts (\textbf{b}), account age (\textbf{c}), and the number of posts per author (\textbf{d}). $\bm{N}=$ \num{82076}.}
    \label{fig:ccdf_comparison}
\end{figure}

Next, we analyze how the characteristics of authors of AI-generated misinformation differ from those of other forms of misinformation (see Fig.~\ref{fig:ccdf_comparison}). 
We observe that AI-generated misleading posts tend to originate from accounts with significantly more followees (\num{7009} vs. \num{5990}), but fewer followers (\num{596041} vs. \num{881884}), and, on average, younger account ages $6.93$ vs. $8.27$ years). These accounts have also posted and interacted with less content throughout their entire lifetime, with a smaller mean post count (\num{55150} vs. \num{73456}). The post count includes original posts, replies, reposts, and quoted posts. 
Further, the number of posts per author is lower for AI-generated posts ($1.44$ vs. $2.43$), which may reflect differences in posting behavior across accounts.
Both two-sided $t$-tests (each $p<0.05$) and Kolmogorov–Smirnov (KS) tests (each $p<0.01$) confirm that these differences between AI-generated and non-AI-generated misinformation are statistically significant. 

\begin{figure}[t]
    \captionsetup{position=top}
    \captionsetup{belowskip=1pt}
    \captionsetup[subfloat]{textfont={sf,normalsize}, skip=2pt, singlelinecheck=false, labelformat=simple, labelfont=bf, justification=centering}
    \centering

    \subfloat[]{%
        \includegraphics[width=0.9\linewidth]{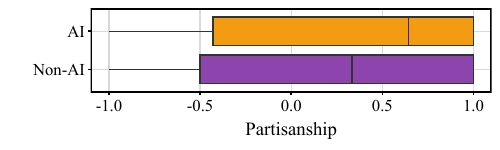}
    }

    \vspace{0.3cm}

    \subfloat[]{%
        \includegraphics[width=0.9\linewidth]{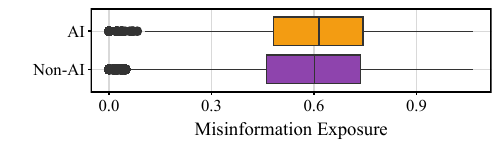}
    }

    \caption{
    Comparison of authors of AI- and non-AI-generated misleading posts ($\bm{N}=$ \num{27985}). 
    Partisanship scores (\textbf{a}) range from \num{-1} (Democrat) to \num{1} (Republican). 
    Misinformation exposure scores (\textbf{b}) range from \num{0} to \num{1}.
    }
    
    \label{fig:partisan_misinfo}
\end{figure}

\textbf{Partisanship and misinformation exposure: } We observe a significantly higher mean partisanship score among authors of AI-generated posts compared to those of non-AI-generated posts (see Fig.~\ref{fig:partisan_misinfo}a), indicating that authors of AI-generated content tend to be more conservative. The partisanship score ranges from \num{-1} (Democrat) to \num{1} (Republican), with higher values indicating greater conservatism \cite{Mosleh.2022}. Specifically, the mean partisanship score for AI-generated posts is \num{0.274}, compared to \num{0.192} for non-AI-generated posts. While both means fall on the conservative side of the scale, the higher value for AI-generated content suggests a notable shift toward more right-leaning users among those sharing such posts. This difference is statistically significant according to a two-sided $t-$test ($t=-4.16$,\, $p<0.001$). Further, authors of AI-generated posts tend to exhibit higher misinformation exposure than those of non-AI-generated posts (see Fig.~\ref{fig:partisan_misinfo}b), with mean scores of \num{0.608} and \num{0.597} respectively. Although the difference is small in magnitude, it is statistically significant ($t=-2.33$,\, $p<0.05$), indicating that authors of AI-generated content are embedded in information environments with slightly higher exposure to misinformation.

\subsection{Virality (RQ~4)}

The posts in our dataset generated more than $168$ billion impressions, were reposted more than $115$ million times, and were liked more than $823$ million times. On average, AI-generated misinformation received \SI{40.19}{\percent} more impressions, \SI{16.90}{\percent} more reposts, and \SI{47.82}{\percent} more likes.

\textbf{Regression model:} 
To better understand the virality of AI-generated vs. other types of misinformation and account for confounding factors, we implement three negative binomial regression models explaining the number of (i) reposts, (ii) likes, and (iii) impressions \cite[\eg,][]{Chuai.2024,Chuai.2024b,Rathje.2021}. The key explanatory variable is $AIGenerated_i$, \ie, whether the post contains AI-generated content ($=1$) or not ($=0$). Consistent with previous work \cite[\eg,][]{Stieglitz.2013,Vosoughi.2018,Drolsbach.2023b}, we control for content characteristics (\ie, \textit{Sentiment}, \textit{Believability}, \textit{Harmfulness}, \textit{Topic}, media type), and account characteristics (\ie, follower count, followee count, account age, verification status). For categorical content variables, we use neutral \textit{Sentiment}, low levels of both \textit{Believability} and \textit{Harmfulness}, and \textit{Other} \textit{Topics} as reference categories

Each model takes the following form: 
{
	\begin{align}
		\log&\left( \mathbb{E}[Y_i \mid \mathbf{X}_i] \right)  = \,\beta_0  + \beta_{1} \, \mathit{AIGenerated}_i  \nonumber  \\
        & + \beta_{2} \,  \mathit{Sentiment}_i  + \beta_{3} \, \mathit{Believability}_i \nonumber \\ 
        & + \beta_{4} \, \mathit{Harmfulness}_i+ \beta_{5} \, \mathit{Topic}_i + \beta_{6} \,  \mathit{MediaType}_i  \nonumber \\ 
        & + \beta_{7} \, \mathit{Followers}_i + \beta_{8} \, \mathit{Followees}_i  \nonumber \\ 
		& + \beta_{9} \, \mathit{Account Age}_i + \beta_{10} \, \mathit{Verified}_i + \gamma_{t} \label{eq:neg_bin_misleading} 
	\end{align} 
}%

where $Y_i$ denotes the outcome variable (reposts, likes, impressions) for post $i$, $\mathbf{X}_i$ is the vector of explanatory variables, and $\beta_0$ the intercept. Furthermore, we include month-year fixed effects $\gamma_{t}$, which allow us to control for variation in exposure windows and platform dynamics over time \cite[\eg,][]{Drolsbach.2023b, Drolsbach.2023a}. In our regression analysis, all continuous variables are $z$-standardized to aid interpretability.

\begin{figure*}[t]
\centering
\includegraphics[width=0.85\textwidth]{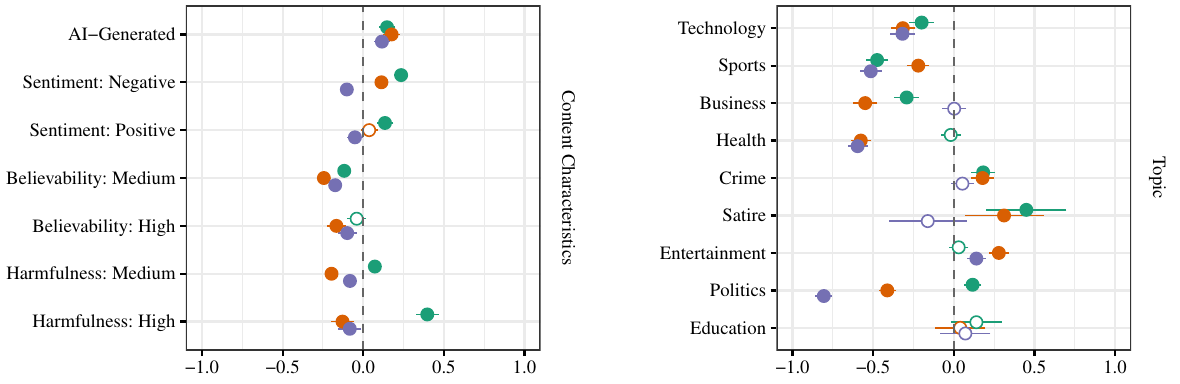}

\includegraphics[width=0.85\textwidth]{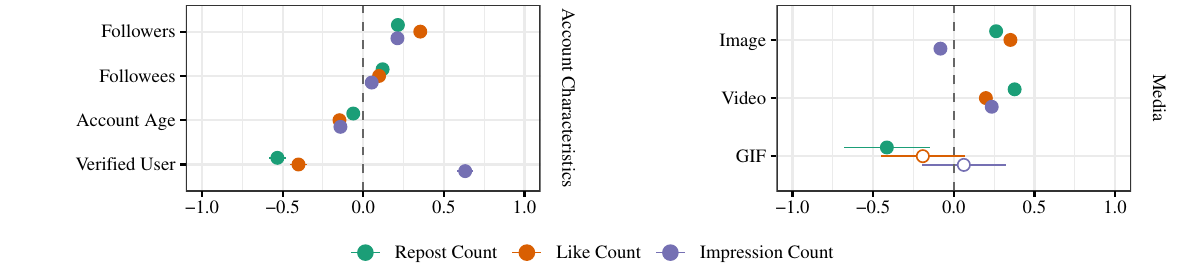}

\caption{Negative binomial regression with the repost count, like count, and impression count as DVs. The circles show standardized coefficient estimates, with filled points denoting statistical significance. The vertical bars represent \SI{99}{\percent} confidence intervals. Month-year fixed effects are included. $\bm{N}=$ \num{82076}.}
\label{fig:model_virality}
\end{figure*}

\textbf{Coefficient estimates: } 
We find that AI-generated misleading posts are substantially more viral than other forms of misinformation. On average, they receive \mbox{$e^{0.148}-1 =$ \SI{15.91}{\percent}} more reposts (coef: $0.148$, \mbox{$p<0.001$}), \SI{19.25}{\percent} more likes (coef: $0.176$, $p<0.001$), and \SI{12.28}{\percent} more impressions (coef: $0.116$, $p<0.001$), compared to non-AI-generated misleading posts (see Fig.~\ref{fig:model_virality}). 

Regarding the control variables, posts with negative \textit{Sentiment} receive \SI{26.50}{\percent} more reposts (coef: $0.235$, \, \mbox{$p<0.001$}) and \SI{11.96}{\percent} more likes (coef: $0.113$, \, \mbox{$p<0.001$}), but \SI{9.67}{\percent} fewer impressions (coef: $-0.102$, \, \mbox{$p<0.001$}), whereas positive \textit{Sentiment} is associated with more reposts (coef: $0.135$, $p<0.001$). Higher levels of \textit{Believability} are associated with reduced engagement. Relative to posts with low \textit{Believability}, posts classified as having medium or high \textit{Believability} receive significantly fewer likes and impressions, while medium-\textit{Believability} posts are also associated with lower repost activity (all $p<0.001$). We found similar patterns for \textit{Harmfulness}, with medium and highly harmful content being associated with reduced likes and impressions, but higher reposts relative to low \textit{Harmfulness} (all $p<0.001$). 
\textit{Topic} effects are heterogeneous, with \textit{Technology} and \textit{Sports} consistently receiving lower engagement across all measures, whereas \textit{Business} is associated with fewer likes and reposts and \textit{Health} with fewer likes and impressions (all $p<0.001$). Conversely, topics such as \textit{Crime}, \textit{Entertainment} and \textit{Satire} are associated with higher engagement, with increased likes across all three topics, more reposts for \textit{Crime} and \textit{Satire}, and more impressions for \textit{Entertainment} (all $p<0.001$). \textit{Politics} is the only topic exhibiting mixed effects, with lower likes and impressions but more reposts (all $p<0.001$), whereas \textit{Education} is not statistically associated with any engagement metric.

Further, we observe that media content plays a significant role in driving user engagement. Compared to posts without media, those containing images receive \SI{30.03}{\percent} more reposts (coef: $0.263$,\, $p<0.001$), while posts with videos receive \SI{45.86}{\percent} more reposts (coef: $0.377$,\, $p<0.001$). Similar patterns are observed across other engagement metrics, except for impressions, where images are associated with a negative significant effect ($p<0.001$). Posts containing images or videos receive significantly higher engagement overall ($p<0.001$), while GIFs show a significant effect only for reposts -- likely due to the small number of posts with GIFs attached in the dataset.

The effects of social influence variables are also statistically significant ($p < 0.001$) and align with findings from prior work \cite[\eg,][]{Drolsbach.2023b}. Higher follower and followee counts are associated with higher engagement, while greater account age is associated with lower engagement. Interestingly, the verification status exhibits a strong positive effect on impression count but a negative effect on both reposts and likes (all $p < 0.001$). This suggests that verified users benefit from greater visibility, potentially due to algorithmic amplification on X, which boosts their content in users' feeds. However, after controlling for other account characteristics (\eg, follower count), this increased exposure does not translate into higher user engagement. 

\textbf{Model checks:}
(1) We verify that variance inflation factors (see SI, Tab.~\ref{tab:vif}), used as indicators of multicollinearity, remain below the commonly accepted threshold of five \cite{Akinwande.2015}.
(2) Community Notes allows contributors to specify why they consider a post misleading, including whether it contains manipulated images or videos (\ie, \textit{misleadingManipulatedMedia}). We therefore repeat our regression analysis while controlling for this indicator. The indicator applies to approximately \SI{13.07}{\percent} of posts in our sample, including \SI{61.52}{\percent} of AI-generated misinformation posts but only \SI{9.24}{\percent} of non-AI-generated misinformation posts. This pattern suggests that AI-generated misinformation is substantially more likely to involve manipulated media, while also indicating that our LLM-based classification captures additional forms of AI-generated misinformation beyond posts explicitly flagged through the Community Notes indicator alone. Importantly, the regression results remain qualitatively unchanged, with AI-generated misinformation continuing to exhibit higher engagement levels than other forms of misleading content (see SI, Tab.~\ref{tab:modelcheck_manipulated_media}).
(3) To ensure that our findings are not driven by uncertain classifications, we restrict the analysis to posts assigned high-confidence labels. The results remain qualitatively unchanged, indicating that classification uncertainty is unlikely to meaningfully affect our conclusions (see SI, Tab.~\ref{tab:modelcheck_highconf}). Moreover, the vast majority of posts in both the AI-generated (\SI{92.81}{\percent}) and non-AI-generated (\SI{99.52}{\percent}) groups receive high-confidence labels, further strengthening confidence in the robustness of our findings.

\section{Discussion}

AI-generated misinformation (\eg, deepfakes) represents a growing challenge in the digital information landscape. Different from traditional forms of misinformation, it leverages advanced AI technologies to create highly realistic yet deceptive content, making detection increasingly difficult  \cite{Feuerriegel.2023, Hancock.2021, Groh.2024, Vaccari.2020}. Despite these risks, research has largely focused on the societal consequences of AI-generated misinformation \cite{Hancock.2021, Vaccari.2020, Dobber.2021} rather than their real-world prevalence. Here, we contribute by conducting a large-scale empirical analysis of AI-generated misinformation circulating on the social media platform X.

\subsection{Implications} 


Our findings highlight the distinct characteristics of AI-generated misinformation and their unique role within the misinformation ecosystem. Compared to other forms of misinformation, AI-generated content is more frequently focused on entertainment and tends to exhibit a more positive sentiment (\textit{RQ~1}), while being perceived as less believable and harmful (\textit{RQ~2}). At the same time, AI-generated misinformation is more likely to originate from smaller user accounts, while authors posting such content are associated with higher levels of partisanship and misinformation exposure (\textit{RQ~3}). Yet, AI-generated misleading posts are significantly more likely to go viral, even after accounting for differences in content and account characteristics (\textit{RQ~4}).

These patterns suggest that the AI-generated nature of content should be treated as a distinct and meaningful factor in misinformation research. Its disproportionate virality points to underlying persuasive properties that may not be captured by existing content-based or intent-based categorizations \cite{Tandoc.2018}. Recent work further suggests that the perception of AI-generated content can vary substantially depending on who or what is depicted in the manipulated content \cite{Sharevski.2025}. Together with our findings, this highlights the importance of incorporating AI-generated content as an explanatory variable in empirical models of misinformation spread and engagement. It also motivates further research into the psychological and perceptual mechanisms that make AI-generated misinformation so compelling. Future research should examine the factors driving its spread (\eg, emotional appeal, visual realism, novelty), the demographics most engaged with such content, and how they shape user interactions differently from traditional forms of misinformation.


For platforms, the high virality of AI-generated misinformation highlights the urgency to develop more effective countermeasures \cite{Feuerriegel.2023}. Traditional strategies (\eg, expert-based fact-checking) often focus on high-profile accounts or recurring misinformation themes \cite{Greene.2025,Chuai.2025}. However, our results demonstrate that smaller accounts are disproportionately responsible for spreading AI-generated misinformation. This indicates that detection strategies must move beyond account size as a primary signal. Here, community-based fact-checking systems -- such as X’s Community Notes -- can be a promising tool by leveraging the collective judgment of users to identify and flag misleading content that may evade both professional fact-checkers and automated systems \cite{Pilarski.2024}. 

At the same time, recent work highlights important challenges in community-based fact-checking systems, including issues related to participation, curation, transparency, and consensus formation \cite{Lloyd.2026}. These systems should therefore complement, rather than replace, broader platform governance efforts. The high virality of AI-generated misinformation further underscores the need for additional platform-level interventions \cite{Feuerriegel.2023,Zhou.2023}, such as AI-based detection systems and strengthened cross-platform coordination. Such measures may become increasingly important given the recent shift toward community-based fact-checking approaches among major social media platforms \cite{Chuai.2026}, including TikTok \cite{TikTok.2025}, YouTube \cite{Youtube.2024}, Facebook, and Instagram \cite{Meta.2025}.

AI-generated misinformation also presents a growing challenge to user trust and safety online \cite{Feuerriegel.2023,Hancock.2021,Goldstein.2023}. As AI-generated content becomes increasingly indistinguishable from authentic content, users face greater challenges in evaluating the authenticity of the information they encounter and share \cite{Groh.2024,Bashardoust.2024,Diel.2024}. This makes media literacy training \cite{Jones.2021,Guess.2020b,Juneja.2022} more essential than ever. Educational initiatives should help users develop skills to critically assess AI-generated misinformation, recognize manipulation techniques, and understand its broader societal implications \cite{Goldstein.2023,Feuerriegel.2023}. Such training is particularly vital for high-stakes domains (\eg, elections \cite{Allcott.2017, Bakshy.2011, McCabe.2024}, and public health \cite{Gallotti.2020, Pennycook.2020b, Solovev.2022b}), where the spread of AI-generated misinformation can have serious real-world consequences \cite{Bar.2023}.

\subsection{Limitations \& Future Research} 

As with any research, our study is not free of limitations. First, our analysis is observational and cannot establish causal relationships. However, our observational approach allows us to uncover robust patterns based on real-world social media data that would be difficult to study in controlled settings. Future work could combine observational analyses with controlled experiments or causal inference approaches to better understand the causal mechanisms underlying the spread of AI-generated misinformation. Second, our data is limited to a single platform (\ie, X), which may constrain the generalizability to other platforms with different user bases or moderation practices. Still, given X's central role in shaping public discourse, it provides a particularly important setting for investigating emerging forms of misinformation. At the same time, future research is needed to examine AI-generated misinformation across additional social media platforms with different moderation systems and user communities. Third, we rely on X’s Community Notes system to identify misinformation, which inherently only captures content that is flagged by users. As such, some AI-generated misleading posts likely go undetected, potentially biasing our view toward more visible or controversial cases. However, this crowd-sourced approach offers a unique advantage in terms of scalability and accuracy, enabling the identification of a wide range of misinformation without relying on manual annotation or automated approaches. Fourth, the accuracy of our findings depends on the reliability of the LLM-based annotations of content characteristics. Yet, prior work demonstrates that LLMs are a reliable tool for large-scale content analysis \cite{Feuerriegel.2025,Rathje.2024}, and we extensively validated our approach through a dedicated user study. Fifth, future studies should examine how fact-checking effectiveness and visibility differ for AI-generated versus traditional misinformation and how corrections influence downstream user behaviors (\eg, sharing). Overall, continued research is needed to understand and mitigate the evolving risks posed by AI-generated misinformation on social media. 

\section{Conclusion}

Despite growing concerns, the characteristics of AI-generated misinformation on social media are poorly understood. Our work addresses this gap by offering a large-scale empirical analysis of AI-generated misinformation circulating on X. Drawing on a dataset of \num{82076} misleading posts identified via X’s Community Notes platform, we show that AI-generated misinformation differs significantly from traditional forms in terms of content attributes, source accounts, and virality. To effectively address this emerging threat, researchers, platforms, and policymakers will need to develop new strategies and countermeasures that account for the unique properties of AI-generated misinformation.

\section{Ethics Statement}
The data collection and analysis follow common standards for ethical research involving digital trace data \cite{Rivers.2014}. Institutional Review Board (IRB) approval was not required because the study relied exclusively on publicly available social media data and did not involve any intervention or interaction with users on the platform. Participants in the user study provided informed consent. We declare no competing interests.


\bibliography{references}

\clearpage
\onecolumn
\appendix
\LARGE
{
\centering
\textbf{Supplementary Materials}
}
\normalsize
\renewcommand\thetable{S\arabic{table}}
\setcounter{table}{0}
\renewcommand\thefigure{S\arabic{figure}}
\setcounter{figure}{0}

\section{Prompt: Identification of AI-Generated Posts}
\label{sec:identification_prompt}

To distinguish AI-generated content from other misleading posts, we used the \texttt{27B} parameter version of \texttt{Gemma 3} \citep{Gemma.2025} to annotate whether the corresponding Community Note indicated that the referenced source post contained AI-generated content. The corresponding prompt is printed below.

\smallskip

{\itshape
``You are a professional annotator specializing on annotating social media content. Your task is to classify social media posts based on Community Notes. The Community Notes have been added for fact-checking purposes to the social media posts, have been found helpful by other users and provide additional context to the social media post. Each post must be annotated along two dimensions: AI-generated, IdentificationConfidence

\smallskip

Your goal is to produce consistent, structured annotations. Follow the definitions and formatting instructions strictly. Return only a valid JSON object with exactly two top-level keys: 'AI-generated' and ``IdentificationConfidence".

\smallskip
\noindent
Definitions and Output Format:

\begin{enumerate}
    \item AI-generated
    \begin{itemize}
    \item Definition: Identify whether you expect the original post to contain AI-generated content based on the community note provided. AI-generated content refers to any form of text, image, audio, video, or other media that is created with the assistance of artificial intelligence technologies. These technologies include machine learning models—such as large language models (LLMs), image generators, or speech synthesis systems—that can mimic human creativity, language, or perception.
    Return only the classification (yes or no).
    \item Output format: "Yes"  // One of: Yes, No
    \end{itemize}
    \item Identification confidence
    \begin{itemize}
    \item  Indicate how confident you are regarding your identification of AI-generated content as provided in 1.
    \item Output format: "Medium"  // One of: Low, Medium, High
\end{itemize}
\end{enumerate}

Return only a valid JSON object using the structure above. Do not include any explanation, comments, or metadata."
}

\section{Prompt: Annotation of Post Characteristics}
\label{sec:annotation_prompt}

We used the \texttt{27B} parameter version of \texttt{Gemma 3} \citep{Gemma.2025} to annotate the source posts with a wide variety of content characteristics. The corresponding prompt is printed below.

\smallskip

{\itshape
``You are a professional content reviewer specializing in analyzing social media posts, including both the post text and attached media (images or video snapshots). All social media posts for comment were identified as misleading by a community-based fact-checking initiative. Your task is to assess the content across the following four dimensions: Believability, Sentiment, Harmfulness, and Topic. Consider both text and visual media in your assessment.
Return only a valid JSON object using the structure and formatting below. Do not include any explanation, comments, or metadata.

\smallskip
\noindent
Definitions and Output Format:

\begin{enumerate}
    \item Believability
\begin{itemize}
    \item Definition: Rate the believability of the social media post (text + attached media). Believability is defined as the quality of something being convincing or credible enough that people accept it as true or plausible. 
    \item Output format: "Believability": "High"  // One of: Low, Medium, High
\end{itemize}

\item Sentiment

\begin{itemize}
    \item Definition: Rate the sentiment of the social media post (text + attached media). Sentiment refers to the underlying emotion or attitude conveyed in a message, reflecting how the author feels about a subject, whether it's positive, negative, or neutral.
    \item Output format: "Sentiment": "Negative"  // One of: Positive, Neutral, Negative
\end{itemize}
\item Harmfulness

\begin{itemize}
    \item Rate the harmfulness of the social media post (text + attached media). Harmfulness is defined as the potential of a (misleading) post to cause real-world damage (such as physical harm, social division, or emotional distress) especially when people believe and act on the misinformation.  
    \item Output format: "Harmfulness": "Medium"  // One of: Low, Medium, High
\end{itemize}

\item Topic

\begin{itemize}
    \item Identify the main topic of the post (text + attached media)
    \item Output format: "Topic": "Politics"  // One of: Technology, Health, Politics, Crime, Business, Entertainment, Sports, Education, Satire, Other"
\end{itemize}
\end{enumerate}
}

\section{Example Annotations}
\label{sec:example_annotation}
\begin{figure*}[h]
    \captionsetup{position=top}
    \captionsetup{belowskip=1pt}
    \captionsetup[subfloat]{textfont={sf,normalsize}, skip=2pt, singlelinecheck=false, labelformat=simple, labelfont=bf, justification=centering}
    \centering

    \subfloat[]{%
        \fbox{%
            \includegraphics[width=0.47\textwidth]{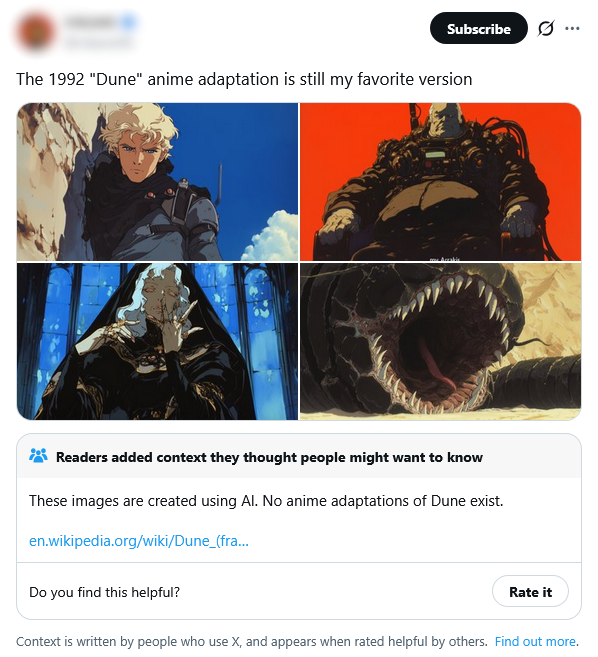}
        }
    }
    \hfill
    \subfloat[]{%
        \fbox{%
            \includegraphics[width=0.47\textwidth]{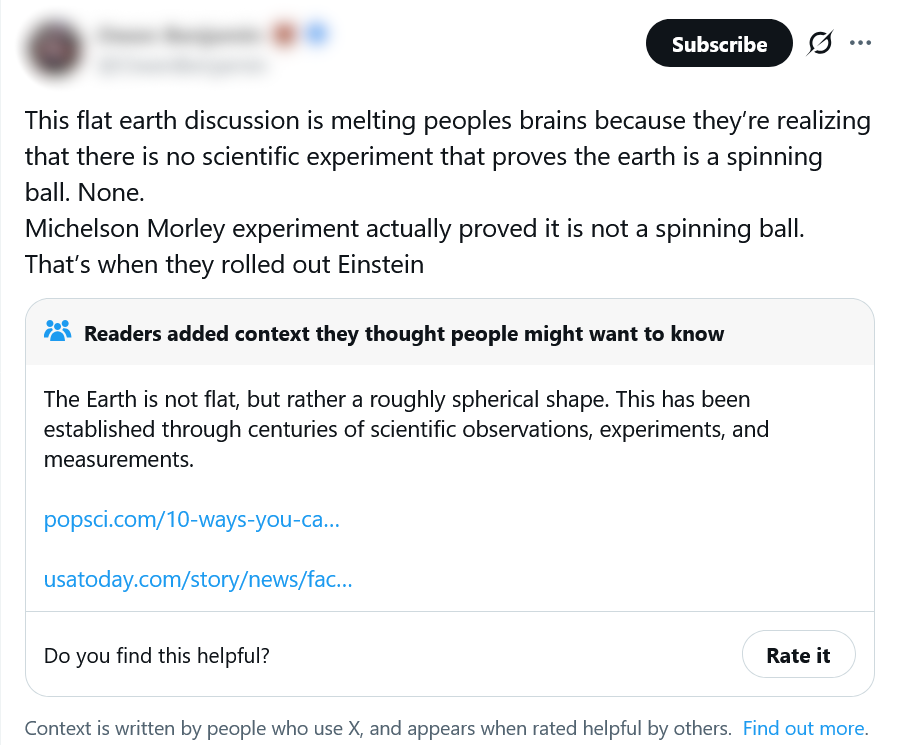}
        }
    }

    \caption{
    Example annotations of (a) AI-generated 
    and (b) non-AI-generated 
    misleading posts, including their corresponding Community Note.
    }
    
    \label{fig:annotated_example}
\end{figure*}

To illustrate the annotation process, Fig.~\ref{fig:annotated_example} presents two example posts together with their corresponding Community Note. In Fig.~\ref{fig:annotated_example}(a), the LLM classifies the content as \textit{AI-generated} (``Yes'', ``High'' confidence) based on the textual explanation provided in the Community Note. Using the post text and associated media (if available), the LLM further assigns the following content labels: \textit{Believability} (``High''), \textit{Sentiment} (``Positive''), \textit{Harmfulness} (``Low''), and \textit{Topic} (``Entertainment''). In contrast, the post in Fig.~\ref{fig:annotated_example}(b) is classified as \textit{AI-generated} (``No'', ``High'' confidence) by the LLM. Based on the post content, the LLM assigns the labels \textit{Believability} (``Low''), \textit{Sentiment} (``Negative''), \textit{Harmfulness} (``Medium''), and \textit{Topic} (``Other'').

\section{Validation}
Table~\ref{tab:user_study_pair} reports strong agreement between the human judgments and the LLM-annotated labels, alongside substantial to almost perfect inter-rater reliability, with mean accuracies ranging from $0.83$ to $0.99$ and Fleiss' $\kappa$ values between $0.64$ and $0.96$.

\begin{table}[H]
    \centering
    \small
    \begin{tabular}{l c c}
        \toprule
        \textbf{Content Characteristic} & \textbf{Average accuracy} & \textbf{Fleiss' $\bm{\kappa}$} \\
        \midrule
        AI-generated     
        & 0.92 & 0.85 \\
        Sentiment   
        & 0.93 & 0.81 \\
        Believability
        & 0.83 & 0.64  \\
        Harmfulness      
        & 0.98 & 0.93  \\
        Topic: Technology              
        & 0.99 & 0.96 \\
        Topic: Sports                      
        & 0.98 & 0.93  \\
        Topic: Business            
        & 0.95 & 0.83 \\
        Topic: Health                      
        & 0.98 & 0.92 \\
        Topic: Crime                 
        & 0.86 & 0.69 \\
        Topic: Satire       
        & 0.91 & 0.75 \\
        Topic: Entertainment
        & 0.98 & 0.95 \\
        Topic: Politics      
        & 0.96 & 0.89 \\
        Topic: Education      
        & 0.96 & 0.89 \\
        \bottomrule
    \end{tabular}
    \caption{Average accuracy across three trained human coders and Fleiss' $\kappa$ for the validation using pairwise comparisons of all annotated content characteristics.}
    \label{tab:user_study_pair}
\end{table}

\section{Summary Statistics}
An overview of the dataset used in this study is shown in Table \ref{tab:data_summary}.

\begin{table}[h]
\centering
\footnotesize
\begin{tabularx}{\textwidth}{
p{2.5cm} 
p{4.5cm} 
S[table-format=4.3, table-number-alignment=center]
S[table-format=4.3, table-number-alignment=center]
S[table-format=4.3, table-number-alignment=center]
S[table-format=4.3, table-number-alignment=center]
S[table-format=4.3, table-number-alignment=center]
S[table-format=4.3, table-number-alignment=center]
}
\toprule
\textbf{Variable} & \textbf{Description} & \multicolumn{2}{c}{\textbf{Overall}} & \multicolumn{2}{c}{\textbf{AI}} & \multicolumn{2}{c}{\textbf{Non-AI}} \\
\cmidrule(lr){3-4} \cmidrule(lr){5-6} \cmidrule(lr){7-8}
 & & \textbf{Mean} & \textbf{SD} & \textbf{Mean} & \textbf{SD} & \textbf{Mean} & \textbf{SD} \\
\midrule
\addlinespace
\multicolumn{8}{l}{\underline{Content Characteristics}} \\
AI-generated & Whether post is identified as AI-generated ($=1$ if true, otherwise~$0$) & 0.073 & 0.261 & 1.000 & 0.000 & 0.000 & 0.000 \\
Sentiment & Sentiment score of the post \mbox{($-1 =$ Negative, $+1 =$ Positive)} & -0.612 & 0.665 & -0.482 & 0.705 & -0.622 &  0.661 \\
Believability: Medium & Medium perceived believability \mbox{($=1$ if Medium)} & 0.315 & 0.464 & 0.169 & 0.374 & 0.326 & 0.469 \\
Believability: High & High perceived believability \mbox{($=1$ if High)} & 0.059 & 0.235 & 0.063 & 0.243 & 0.059 & 0.235 \\
Harmfulness: Medium & Medium perceived harmfulness \mbox{($=1$ if Medium)} & 0.662 & 0.473 & 0.572 & 0.495 & 0.669 & 0.471 \\
Harmfulness: High & High perceived harmfulness \mbox{($=1$ if High)} & 0.054 & 0.226 & 0.047 & 0.212 & 0.055 & 0.227 \\
Topic & Topic category of the post 
& {--} & {--} & {--} & {--} & {--} & {--} \\
Media Type & Type of media included in the post 
& {--} & {--} & {--} & {--} & {--} & {--} \\
\addlinespace
\multicolumn{8}{l}{\underline{Account Characteristics}} \\
Followers & Followers (in 1000s) & 860.904 & 6972.571 & 596.041 & 6096.824 & 881.884 & 7036.891 \\
Followees & Accounts followed (in 1000s) & 6.064 & 24.132 & 7.009 & 32.137 & 5.990 & 23.380 \\
Account Age & Account age (in years) & 8.177 & 5.074 & 6.939 & 4.801 & 8.275 & 5.082 \\
Verified User & Verified account ($=1$~if~true) & 0.078 & 0.268 & 0.043 & 0.202 & 0.081 & 0.272 \\
Post Count & Posts, Reposts, Quotes \& Replies (in 1000s) & 72.112 & 139.208 & 55.150 & 90.698 & 73.456 & 142.260 \\
Partisanship & Political leaning ($-1 =$ Democrat, $+1 =$ Republican) & 0.197 & 0.771 & 0.274 & 0.779 & 0.192 & 0.770 \\
Misinformation exposure & Exposure to misinformation $[0,1]$ & 0.584 & 0.168 & 0.595 & 0.165 & 0.584 & 0.168 \\

\addlinespace
\multicolumn{8}{l}{\underline{Virality}} \\
Repost Count & Number of reposts (in 1000s) & 1.404 & 3.215 & 1.621 & 3.322 & 1.387 & 3.206 \\
Like Count & Number of likes (in 1000s) & 10.028 & 24.691 & 14.320 & 29.530 & 9.688 & 24.234 \\
Impression Count & Number of impressions (in 1000s) & 2057.968 & 6655.169 & 2802.440 & 10080.099 & 1998.999 & 6301.202 \\

\bottomrule
\end{tabularx}
\caption{Variable Definitions and Summary Statistics.\label{tab:data_summary}}
\end{table}

\section{Model Checks}
Table \ref{tab:vif} reports the variance inflation factors for all explanatory variables in our regression models. All VIF values remained substantially below the commonly accepted threshold of five, suggesting that multicollinearity was not an issue in our analysis. Table \ref{tab:modelcheck_manipulated_media} presents a regression model including \textit{misleadingManipulatedMedia} as an additional predictor. Table \ref{tab:modelcheck_highconf} reports a regression model restricted to posts with high-confidence LLM annotation of \textit{AI-generated}. Across both regression models, the effect of \textit{AI-generated} remained positive and statistically significant.

\begin{table}[H]
    \centering
    \footnotesize
    \begin{tabular}{l c c c}
        \toprule
        & \textbf{Repost Count} & \textbf{Like Count} & \textbf{Impression Count} \\
        \midrule
        AI-generated     
        & 1.04 & 1.04 & 1.04 \\
        Sentiment   
        & 1.53 & 1.53 & 1.53 \\
        Believability
        & 1.21 & 1.21 & 1.21  \\
        Harmfulness      
        & 2.06 & 2.06 & 2.06  \\
        Topic          
        & 1.94 & 1.94 & 1.94 \\
        Media                      
        & 1.12 & 1.12 & 1.12  \\
        Follower            
        & 1.02 & 1.02 & 1.02 \\
        Followees                      
        & 1.01 & 1.01 & 1.01 \\
        Account Age                 
        & 1.11 & 1.11 & 1.11 \\
        Verified       
        & 1.06 & 1.06 & 1.06 \\
        \bottomrule
    \end{tabular}
    \caption{Variance inflation factors for all three negative binomial regression models.}
    \label{tab:vif}
\end{table}
\clearpage

\begin{table}[H]
\centering
\small
\renewcommand{\arraystretch}{1.5}

\begin{tabular*}{\textwidth}{@{\extracolsep{\fill}}l *{3}{S[table-format=-1.3,table-space-text-post=***,table-align-text-post=false]@{\,}l}}
\toprule
& \multicolumn{2}{c}{\textbf{Repost Count}}
& \multicolumn{2}{c}{\textbf{Like Count}}
& \multicolumn{2}{c}{\textbf{Impression Count}} \\
\midrule
 & {Coef.} & {Std. Error} & {Coef.} & {Std. Error} & {Coef.} & {Std. Error} \\
 \cmidrule(lr){2-3} \cmidrule(lr){4-5} \cmidrule(lr){6-7}
\underline{Content Characteristics} \\

AI-generated
& 0.183***
& {(0.022)}
& 0.214***
& {(0.021)}
& 0.182***
& {(0.021)} \\

Misleading: Manipulated Media
& -0.067***
& {(0.017)}
& -0.076***
& {(0.017)}
& -0.136***
& {(0.016)} \\

Sentiment: Negative
& 0.236***
& {(0.016)}
& 0.113***
& {(0.016)}
& -0.102***
& {(0.016)} \\

Sentiment: Positive
& 0.136***
& {(0.020)}
& 0.037
& {(0.020)}
& -0.048*
& {(0.020)} \\

Believability: Medium
& -0.121***
& {(0.012)}
& -0.247***
& {(0.012)}
& -0.180***
& {(0.012)} \\

Believability: High
& -0.042
& {(0.023)}
& -0.168***
& {(0.023)}
& -0.106***
& {(0.022)} \\

Harmfulness: Medium
& 0.070***
& {(0.016)}
& -0.198***
& {(0.016)}
& -0.085***
& {(0.016)} \\

Harmfulness: High
& 0.395***
& {(0.028)}
& -0.130***
& {(0.028)}
& -0.085**
& {(0.027)} \\

\addlinespace
\underline{Topic} \\

Technology
& -0.203***
& {(0.030)}
& -0.319***
& {(0.029)}
& -0.324***
& {(0.029)} \\

Sports
& -0.477***
& {(0.026)}
& -0.222***
& {(0.026)}
& -0.515***
& {(0.026)} \\

Business
& -0.296***
& {(0.030)}
& -0.554***
& {(0.029)}
& 0.003
& {(0.029)} \\

Health
& -0.017
& {(0.025)}
& -0.578***
& {(0.024)}
& -0.603***
& {(0.024)} \\

Crime
& 0.179***
& {(0.029)}
& 0.174***
& {(0.029)}
& 0.047
& {(0.028)} \\

Satire
& 0.448***
& {(0.096)}
& 0.310**
& {(0.095)}
& -0.162
& {(0.094)} \\

Entertainment
& 0.031
& {(0.024)}
& 0.281***
& {(0.023)}
& 0.142***
& {(0.023)} \\

Politics
& 0.115***
& {(0.021)}
& -0.415***
& {(0.020)}
& -0.810***
& {(0.020)} \\

Education
& 0.134*
& {(0.061)}
& 0.033
& {(0.061)}
& 0.063
& {(0.060)} \\

\addlinespace
\underline{Media} \\

Image
& 0.267***
& {(0.013)}
& 0.358***
& {(0.013)}
& -0.075***
& {(0.013)} \\

Video
& 0.381***
& {(0.014)}
& 0.204***
& {(0.014)}
& 0.241***
& {(0.014)} \\

GIF
& -0.417***
& {(0.103)}
& -0.194
& {(0.102)}
& 0.057
& {(0.101)} \\

\addlinespace
\underline{Account Characteristics} \\

Followers
& 0.214***
& {(0.006)}
& 0.352***
& {(0.005)}
& 0.210***
& {(0.005)} \\

Followees
& 0.120***
& {(0.005)}
& 0.098***
& {(0.005)}
& 0.053***
& {(0.005)} \\

Account Age
& -0.063***
& {(0.005)}
& -0.148***
& {(0.005)}
& -0.144***
& {(0.005)} \\

Verified User
& -0.535***
& {(0.020)}
& -0.404***
& {(0.019)}
& 0.635***
& {(0.019)} \\

\addlinespace
\underline{Additional Controls} \\

Month-Year Fixed Effects
& \multicolumn{2}{c}{Included}
& \multicolumn{2}{c}{Included}
& \multicolumn{2}{c}{Included} \\

\midrule

AIC
& \multicolumn{2}{c}{1,294,884}
& \multicolumn{2}{c}{1,612,454}
& \multicolumn{2}{c}{2,483,154} \\

Observations
& \multicolumn{2}{c}{82,076}
& \multicolumn{2}{c}{82,076}
& \multicolumn{2}{c}{82,076} \\

\bottomrule

\multicolumn{7}{l}{
\small{Significance levels: $^{***}p<0.001$; $^{**}p<0.01$; $^{*}p<0.05$}
} \\

\end{tabular*}
\caption{Negative binomial regression model including \textit{misleadingManipulatedMedia} as an additional predictor.} \label{tab:modelcheck_manipulated_media}

\end{table}

\clearpage

\begin{table}[H]
\centering
\small
\renewcommand{\arraystretch}{1.5}
\begin{tabular*}{\textwidth}{@{\extracolsep{\fill}}l *{3}{S[table-format=-1.3,table-space-text-post=***,table-align-text-post=false]@{\,}l}}
\toprule
& \multicolumn{2}{c}{\textbf{Repost Count}}
& \multicolumn{2}{c}{\textbf{Like Count}}
& \multicolumn{2}{c}{\textbf{Impression Count}} \\
\midrule
 & {Coef.} & {Std. Error} & {Coef.} & {Std. Error} & {Coef.} & {Std. Error} \\
 \cmidrule(lr){2-3} \cmidrule(lr){4-5} \cmidrule(lr){6-7}
\underline{Content Characteristics} \\

AI-generated
& 0.147***
& {(0.021)}
& 0.176***
& {(0.020)}
& 0.093***
& {(0.020)} \\

Sentiment: Negative
& 0.235***
& {(0.016)}
& 0.112***
& {(0.016)}
& -0.100***
& {(0.016)} \\

Sentiment: Positive
& 0.137***
& {(0.020)}
& 0.037
& {(0.020)}
& -0.058**
& {(0.020)} \\

Believability: Medium
& -0.119***
& {(0.012)}
& -0.245***
& {(0.012)}
& -0.172***
& {(0.012)} \\

Believability: High
& -0.042
& {(0.023)}
& -0.169***
& {(0.023)}
& -0.097***
& {(0.023)} \\

Harmfulness: Medium
& 0.071***
& {(0.016)}
& -0.199***
& {(0.016)}
& -0.088***
& {(0.016)} \\

Harmfulness: High
& 0.397***
& {(0.028)}
& -0.130***
& {(0.028)}
& -0.085**
& {(0.028)} \\

\addlinespace
\underline{Topic} \\

Technology
& -0.189***
& {(0.030)}
& -0.311***
& {(0.029)}
& -0.330***
& {(0.029)} \\

Sports
& -0.470***
& {(0.027)}
& -0.213***
& {(0.026)}
& -0.515***
& {(0.026)} \\

Business
& -0.287***
& {(0.030)}
& -0.542***
& {(0.029)}
& 0.001
& {(0.029)} \\

Health
& -0.018
& {(0.025)}
& -0.575***
& {(0.025)}
& -0.601***
& {(0.024)} \\

Crime
& 0.191***
& {(0.029)}
& 0.186***
& {(0.029)}
& 0.051
& {(0.028)} \\

Satire
& 0.461***
& {(0.096)}
& 0.324***
& {(0.096)}
& -0.154
& {(0.095)} \\

Entertainment
& 0.039
& {(0.024)}
& 0.288***
& {(0.023)}
& 0.133***
& {(0.023)} \\

Politics
& 0.124***
& {(0.021)}
& -0.405***
& {(0.020)}
& -0.805***
& {(0.020)} \\

Education
& 0.153*
& {(0.062)}
& 0.056
& {(0.061)}
& 0.076
& {(0.061)} \\

\addlinespace
\underline{Media} \\

Image
& 0.258***
& {(0.013)}
& 0.345***
& {(0.013)}
& -0.078***
& {(0.013)} \\

Video
& 0.372***
& {(0.014)}
& 0.194***
& {(0.014)}
& 0.242***
& {(0.014)} \\

GIF
& -0.428***
& {(0.103)}
& -0.209*
& {(0.102)}
& 0.075
& {(0.101)} \\

\addlinespace
\underline{Account Characteristics} \\

Followers
& 0.218***
& {(0.006)}
& 0.362***
& {(0.006)}
& 0.218***
& {(0.005)} \\

Followees
& 0.121***
& {(0.005)}
& 0.100***
& {(0.005)}
& 0.053***
& {(0.005)} \\

Account Age
& -0.062***
& {(0.005)}
& -0.148***
& {(0.005)}
& -0.141***
& {(0.005)} \\

Verified User
& -0.533***
& {(0.020)}
& -0.403***
& {(0.020)}
& 0.632***
& {(0.019)} \\

\addlinespace
\underline{Additional Controls} \\

Month-Year Fixed Effects
& \multicolumn{2}{c}{Included}
& \multicolumn{2}{c}{Included}
& \multicolumn{2}{c}{Included} \\

\midrule

AIC
& \multicolumn{2}{c}{1,282,610}
& \multicolumn{2}{c}{1,596,853}
& \multicolumn{2}{c}{2,458,440} \\

Observations
& \multicolumn{2}{c}{81,282}
& \multicolumn{2}{c}{81,282}
& \multicolumn{2}{c}{81,282} \\

\bottomrule

\multicolumn{7}{l}{
\small{Significance levels: $^{***}p<0.001$; $^{**}p<0.01$; $^{*}p<0.05$}
} \\

\end{tabular*}
\caption{Negative binomial regression model restricted to high-confidence LLM classifications.} \label{tab:modelcheck_highconf}

\end{table}

\end{document}